# Coherent control of magnetization precession in ferromagnetic semiconductor (Ga,Mn)As

E. Rozkotová, P. Němec[a)], N. Tesařová, and P. Malý
*Faculty of Mathematics and Physics, Charles University in Prague, Ke Karlovu 3,
121 16 Prague 2, Czech Republic*

V. Novák, K. Olejník, M. Cukr, and T. Jungwirth
*Institute of Physics ASCR v.v.i., Cukrovarnická 10, 162 53 Prague, Czech Republic*

We report single-color, time resolved magneto-optical measurements in ferromagnetic semiconductor (Ga,Mn)As. We demonstrate coherent optical control of the magnetization precession by applying two successive ultrashort laser pulses. The magnetic field and temperature dependent experiments reveal the collective Mn-moment nature of the oscillatory part of the time-dependent Kerr rotation, as well as contributions to the magneto-optical signal that are not connected with the magnetization dynamics.

The lasting demand for increased speed of writing and retrieving magnetically stored information in computer hard drives stimulated an intense research of ultrafast magnetization dynamics. In particular, the ultrafast control of magnetization by laser pulses has gained a significant attention recently [1]. Initially, the research was focused mainly on ferromagnetic metals and half-metals where the impact of ultrashort (subpicosecond) laser pulses can lead to demagnetization [2], magnetization rotation [3] or even to modification of magnetic structure [4]. More recently, also the investigation of diluted magnetic semiconductors (DMSs) has started. The most intensively studied example of DMSs is (Ga,Mn)As where the ferromagnetic coupling between Mn local-moments is meadiated by spin-polarized valence band holes [5]. In the last few years, not only the ultrafast demagnetization [6] but also the ultrafast enhancement of ferromagnetism [7], the complete reversal of magnetic hysteresis loop [8] and the laser-induced precession of magnetization [9-11] were demonstrated in (Ga,Mn)As.

Ultrashort laser pulses can be also used for *coherent* control of the spin precession. In this experiment the sample is excited by pairs of pump pulses. Each pump pulse generates a transient precession of magnetization vector and their temporal separation (i.e., the mutual phase difference between the corresponding magnetization precessions) determines if they superimpose constructively or destructively. Coherent control of magnetization was demonstrated in various types of magnetic materials – ferrimagnetic garnet [1], antiferromagnetic orthoferrites [1], half-metalic ferromagnetic $CrO_2$ [12], and paramagnetic (Cd,Mn)Te [13].

In this paper, we report on the coherent control of magnetization precession in ferromagnetic (Ga,Mn)As. The experiments were performed on an annealed 50 nm thick ferromagnetic (Ga,Mn)As film with nominal *Mn* doping of 7% and the Curie temperature $T_C \approx 160$ K, which was grown by the low temperature molecular beam epitaxy (LT-MBE) on a GaAs(001) substrate. The sample exhibits in-plane easy axis behavior typical for stressed (Ga,Mn)As layers grown on GaAs substrates. The external magnetic field (generated by an

---

[a)] Electronic mail: nemec@karlov.mff.cuni.cz

electromagnet) was applied in the sample plane along the [010] crystallographic direction. In the experiment, the sample was always cooled with a magnetic field of 700 mT. The photoinduced magnetization dynamics was studied by the time-resolved magneto-optical Kerr effect (TR-MOKE) technique [14] using a femtosecond titanium sapphire laser (Tsunami, Spectra Physics). Laser pulses with the time width of 80 fs and the repetition rate of 82 MHz were tuned to 1.61 eV. The energy fluence of the pump laser pulses was about 40 μJ.cm$^{-2}$ and the probe pulses were always at least 10 times weaker. The angles of incidence (measured from the normal of the film surface) of the pump "1", pump "2" and probe pulses were 2°, 25° and 8°, respectively. The polarization of the pump pulses was either circular or linear, while the probe pulses were linearly polarized (along the [100] crystallographic direction in the sample). The rotation angle of the polarization plane of the reflected probe pulses was obtained by taking the *difference* of signals measured by detectors in an optical bridge detection system [14]. Simultaneously, we measured also the *sum* of signals from the detectors, which corresponds to the change of the sample reflectivity induced by the pump pulse [11].

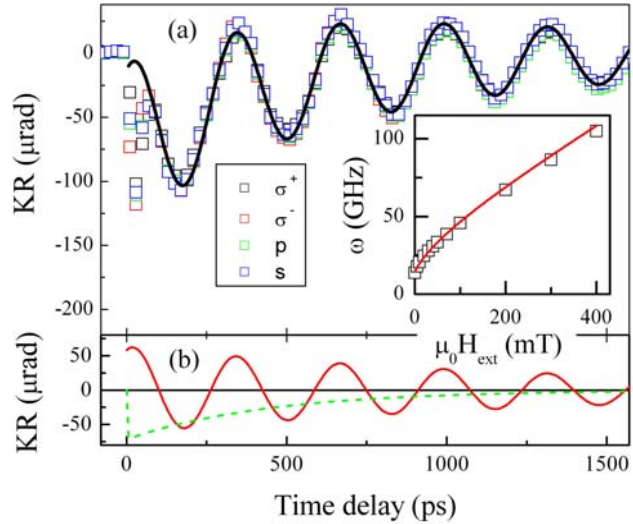

Fig. 1. (a) Dynamics of KR measured for circularly ($\sigma^+$ and $\sigma^-$) and linearly (*p* and *s*) polarized pump pulses (points); sample temperature 8 K, external magnetic field $H_{ext}$ = 10 mT. The solid line is a fit by a sum of the exponentially damped sine harmonic oscillation ($A_1 \exp(-t/\tau_D)\sin(\omega t + \varphi)$; solid line in part (b)) and the pulse-like KR signal ($C_1 [1-\exp(-t/\tau_1)]\exp(-t/\tau_2)$; dashed line in part (b)); the parameters of the fit are: $A_1$ = 63 μrad, $\tau_D$ = 1.5 ns, $\omega$ = 19.4 GHz, $\varphi$ = 68°, $C_1$ = - 71 μrad, $\tau_1$ = 0.5 ps, $\tau_2$ = 440 ps. Inset: Dependence of the angular frequency $\omega$ on $H_{ext}$ (points). The line is the fit by $\omega = \gamma \sqrt{(H_{ext} + H_{4\parallel} + H_a)(H_{ext} + H_{4\parallel})}$, where $H_a = 4\pi M - H_{2\perp} + H_{2\parallel}/2$ (see Ref. 16 for the derivation of this equation and for the used notation); the parameters of the fit are: $\gamma$ = 2.2 x 10$^5$ mA$^{-1}$s$^{-1}$ (g-factor of 2), $\mu_0 H_{4\parallel}$ = 13 mT, $\mu_0 H_a$ = 500 mT.

In ferromagnetic (Ga,Mn)As the impact of strong femtosecond laser pulses induces a change of the magnetic anisotropy that leads to a precession of magnetization (even when no external magnetic field is applied), which can be detected as an oscillatory temporal trace of the transient angle of Kerr rotation (KR) [9-11]. In Fig. 1(a), we show the KR signals measured for different polarizations of pump pulses at 8 K (points). The data are very similar to those that we reported previously for slightly different samples [11, 15]. In particular, the signal is nearly independent of the polarization of the pump pulses. This implies that the magnetic anisotropy change is not induced by the spin of carriers photogenerated in the sample by absorption of circularly polarized photons. Instead, the change of hole concentration and/or the local temperature increase are probably responsible for this effect [9-11]. The measured KR traces can be fitted well (see solid line in Fig. 1(a)) by an exponentially damped sine harmonic oscillation superimposed on a pulse-like function, as described by Eq. (1) in [11] and illustrated in Fig. 1(b). It is also worth noting that the data can be fitted well only for the time delays larger than ≈ 70 ps. In particular, the presence of a sharp negative peak is not reproduced by the fit; we will discuss this point later. The angular frequency $\omega$ of the oscillations increases with the external magnetic field applied in the sample plane (points in the inset of Fig. 1) as predicted by the classical gyromagnetic theory [16]. The most important point in the present discussion is that the measured data can be fitted well by the expected dependence [16] with the value of the gyromagnetic constant $\gamma$ characteristic for $Mn^{2+}$ spins (see the inset of Fig. 1). This fact, together with the disappearance of oscillations for sample temperatures above $T_C$ (see Fig. 3(c) in [11]), confirm that the oscillatory KR signal is connected with the precession of the *ferromagnetically coupled Mn* spins in (Ga,Mn)As.

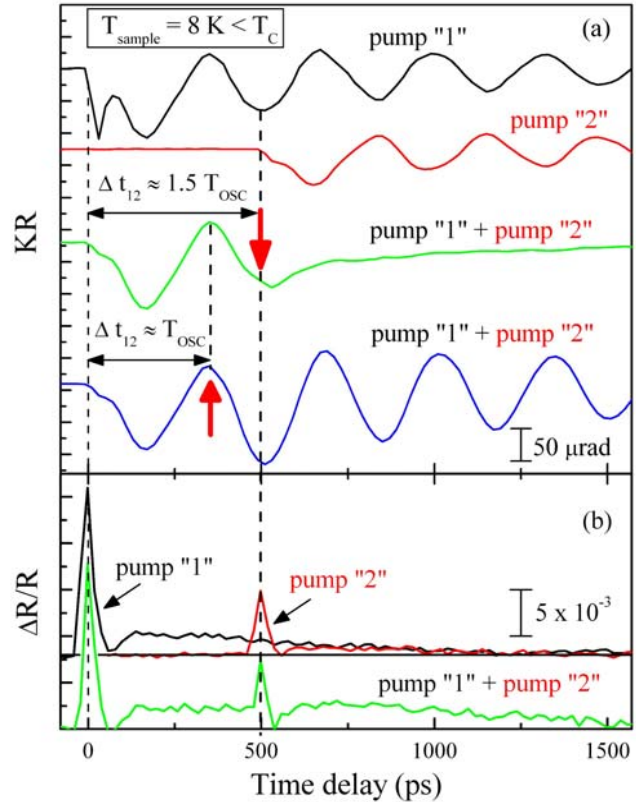

Fig. 2. (a) Double-pump KR experiment below the Curie temperature; sample temperature 8 K, external magnetic field 10 mT, *s* linearly polarized laser pulses. Each pump pulse alone triggers precession of magnetization with a precessional period $T_{OSC}$, as shown by the top and the second from top traces for pump pulses "1" and "2", respectively. If pump pulse "2" excites the sample 500 ps ≈ 1.5 $T_{OSC}$ after the pulse "1" (as shown by the vertical arrow), the magnetization precession is stopped. If the time delay between the pulses is 340 ps ≈ $T_{OSC}$, the magnetization precession is amplified (the bottom trace). The curves are vertically shifted for clarity. (b) Dynamics of the reflectivity change, which were measured simultaneously with the dynamics of KR shown in (a). The horizontal line corresponds to $\Delta R = 0$; the bottom curve was vertically shifted for clarity.

The coherent control of magnetization precession is illustrated in Fig. 2 (a). The top trace in Fig. 2 (a) shows how a pump pulse "1" arriving at zero time delay triggers precession of magnetization, as described in the previous paragraph. The time delayed pump pulse "2" *alone* has exactly the same influence on the magnetization, as seen on the second from top trace in Fig. 2 (a). When *both* pulses excite the sample the situation changes significantly. If a time delay between the pulses "1" and "2" ($\Delta t_{12}$) is equal to an odd multiple of a half of the precessional period $T_{OSC}$, the magnetization precession is suppressed. On the contrary, if the second pump pulse arrives at an integer number of full periods, the subsequent precession is enhanced. The results of the analysis of the measured data are shown in Fig. 3. If only pump "1" excites the sample the precession amplitude $A_1$ is slowly damped in time (black solid line in Fig. 3). When both pump pulses excite the sample the resulting precession amplitude is very sensitive to the value of $\Delta t_{12}$ (solid points). In the vicinity of $\Delta t_{12} = 1.5\ T_{OSC}$, which corresponds to the mutual phase difference between the oscillations $\Delta\Phi = 3\pi$, this dependence can be approximated by the function $1+\cos\Delta\Phi$ (dashed line in Fig. 3), as expected for the sum of two mutually phase shifted harmonic functions. On the other hand, the amplitude of the pulse-like function $C$ (open points in Fig. 3) is only weakly dependent on $\Delta t_{12}$. We should also mention that for the efficient coherent control of the magnetization dynamics, the amplitudes of the precessions induced by pump pulse "1" ($A_1$) and "2" ($A_2$) had to be the same in a moment of the impact of pump pulse "2". Consequently, due to the damping of $A_1$ in time, the intensity of pump pulse "2" had to be lower than that of pump pulse "1".

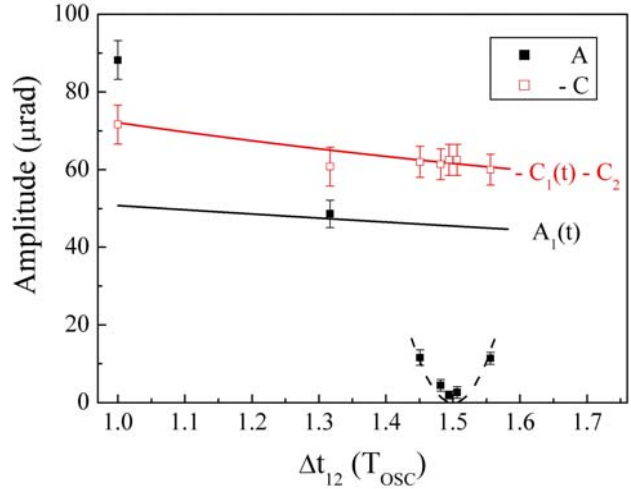

Fig. 3. Dependence of $A$ and $-C$ on a time delay $\Delta t_{12}$ between the pump pulses "1" and "2" in the double-pump KR experiment (points); the time delay is expressed in multiples of the precession period $T_{OSC} = 323$ ps. The black solid line shows the evolution of $A_1$ in time and the red line corresponds to $-C = -C_1\exp(-\Delta t_{12}T_{OSC}/\tau_2) - C_2$, where the values of all parameters were obtained from the single-pump experiments with pump pulse "1" (see Fig. 1) and "2" ($C_2 = -38$ μrad). The dashed line depicts the function $1 + \cos(2\pi\Delta t_{12})$.

In general, the measured magneto-optical KR signal consists of two components – "magnetic" and "non-magnetic" (see Eq. (2) in [17]). The former reflects the pump-induced change of the sample magnetization and the latter is connected with the pump-induced change of the electronic properties of the sample [17]. Recently, we have showed that the oscillatory part of the signal is very similar in the Kerr rotation and ellipticity measurements (see Fig. 1 in [15]), which confirm its "magnetic" origin. In Fig. 2 (b) we show the measured dynamics of the sample reflectivity change $\Delta R/R$, which monitor the modification of the complex index of refraction induced in the sample by the carriers photoinjected by the pump pulses. The dynamics of $\Delta R/R$ induced by the pump pulses "1" and "2" is very similar. The only difference is in the signal magnitude that is a consequence of the lower intensity of the pump pulse "2". When both pump pulses excite the sample, the signals add up (see bottom trace in Fig. 2 (b)) and show no significant dependence on $\Delta t_{12}$ (not shown here). The similar insensitivity of the magnitude of $\Delta R/R$ (at time delay ≥ 150 ps after the impact of the pump pulse "2") and the amplitude of the pulse-like function *C* (see Fig. 3) on the value of $\Delta t_{12}$ suggests that the pulse-like KR signal is of "non-magnetic" origin or that it is connected with the pump pulse-induced temperature increase of the lattice. To further test this hypothesis we repeated the double-pump experiment at sample temperature *above* its Curie temperature - see Fig. 4. The absence of the oscillations is in agreement with our previous conclusion that *only* the oscillatory signal is directly connected with the ferromagnetic order in the sample [11, 15]. When both pump pulses excite the sample above $T_c$, the signals add up and show no dependence on $\Delta t_{12}$. We also found out that the measured signal above $T_c$ does not change significantly with the external magnetic field (up to 700 mT). This confirms that the pulse-like part of KR signal above $T_c$ is connected with the pump-induced modification of the complex index of refraction of the sample (presumably by the carriers trapped in defect states in this LT-MBE grown material). This assignment is also supported by the observed reduction of the magnitude of the KR signal induced above $T_C$ by the pump pulse "2" in the double-pump experiment compared to that induced by the pump pulse "2" in the single-pump experiment (see Fig. 4). The available defect states are probably already partially filled by carriers photoexcited by the pump pulse "1" and, therefore, the pump pulse "2" can induce only a smaller change of the complex index of refraction. Similarly, also the sharp negative peak observed in the KR signal at low temperature (see Fig. 1(a)) is probably of "non-magnetic" origin as indicated by its sensitivity to the angle of incidence of the pump pulses on the sample (it is rather strong after excitation by pump pulse "1" and much weaker after excitation by pump pulse "2" – see Fig. 2). Moreover, this peak is also strongly suppressed in the double-pump experiments (cf. the top and the bottom traces in Fig. 2 (a)), which indicates its sensitivity to the time-spacing between the adjacent laser pulses coming from the laser (due to the laser repetition rate of 82 MHz the time-spacing between the pulses is ≈ 12 ns in the single-pump experiment).

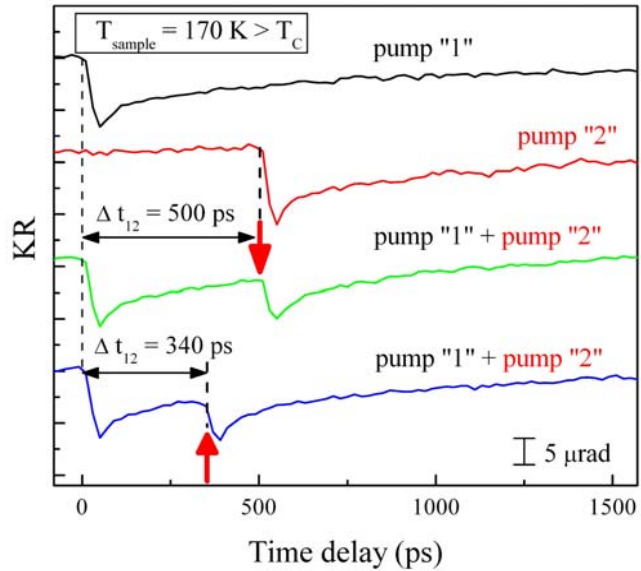

Fig. 4. Double-pump KR experiment above the Curie temperature; sample temperature 170 K, external magnetic field 10 mT, *s* linearly polarized laser pulses. The notation is the same as in Fig. 2. Note a reduction of the signal amplitude with the sample temperature.

In conclusion, we performed double-pump TR-MOKE measurements in (Ga,Mn)As. Our data clearly show that femtosecond laser pulses can be used to coherently control magnetization dynamics in ferromagnetic semiconductors. In addition to this, the analysis of the experiments identified various "non-magnetic" contributions to the measured KR signals that should be carefully considered in any quantitative studies of the ultrafast pulse-induced magnetization dynamics.

This work was supported by Ministry of Education of the Czech Republic in the framework of the research centre L510, the research plans MSM0021620834 and AV0Z1010052, by the Grant Agency of the Charles University in Prague under Grant No. 252445, and by the Grant Agency of Academy of Sciences of the Czech Republic Grants FON/06/E001, FON/06/E002, and KAN400100652.

*Note added*: After the preparation of the manuscript was completed, similar experimental results of a double-pump KR experiment (below $T_C$) have been presented by Munakata at the PASPS V conference [18].